# Theoretical prediction of superconductivity in monolayer h-BN doped with alkaline-earth metals (Ca, Sr, Ba)

Nao. H. Shimada[1], Emi Minamitani*[1,2], Satoshi Watanabe[1]

[1]*Department of Materials Engineering, The University of Tokyo, Japan*

[2]*Institute for Molecular Science, Japan.*

eminamitani@ims.ac.jp

**Abstract**

We investigated the possibility of superconductivity in monolayer hexagonal boron nitride (h-BN) doped using each group-1 (Li, Na, K) and group-2 (Be, Mg, Ca, Sr, Ba) atom via ab-initio calculations. Consequently, we reveal that Sr- and Ba-doped monolayer h-BN and Ca-doped monolayer h-BN with 3.5% tensile strain are energetically stable and become superconductors with Tc values of 5.83 K, 1.53 K, and 12.8 K, respectively, which are considerably higher than those of Ca-, Sr-, and Ba-doped graphene. In addition, the momentum-resolved electron–phonon coupling (EPC) constant shows that the scattering among intrinsic $\pi^*$ electrons around the $\Gamma$ point dominates Tc. The scattering process is mediated by the low-energy vibration of the adsorbate. Moreover, compared with graphene, the stronger adsorbate–substrate interaction and lower symmetry in h-BN are critical for enhancing EPC in doped h-BN.

**Introduction**

Two-dimensional (2D) materials have attracted considerable attention as the candidates for materials with novel physical properties[1]. The superconductivity in 2D materials is expected to be different than that in three-dimensional (3D) bulk materials and could have a high potential for hosting unique phenomena such as quantum phase transition. Therefore, finding a new 2D superconductor is one of the important topics in this field. Although several types of 2D superconductors have been experimentally and/or theoretically investigated, most of them are metallic/semimetallic materials, such as metal atomic layers on semiconductor surfaces[2–4], and layered materials such as FeSe[5,6], NbSe$_2$[7], and electron-doped graphene[8–15]. The other class of 2D superconductors is doped 2D semiconductors or insulators. In doped 2D semiconductors, the superconductivity induced via gate doping by using ionic liquid was recently reported in MoS$_2$[16–18] and ZrNCl[19]. The other route to realizing superconductivity is via chemical doping using alkali or alkaline-earth metals. Although the tunability of the doping amount in chemical doping is lower than that in gate doping, the former offers additional electron–phonon interaction processes due to the low-energy phonon modes and electronic states induced by the dopant, thereby increasing the superconducting transition temperature. In addition, the superconductivity in chemically doped MoS$_2$[20] and black phosphorus[21] has been reported. However, the investigation on the possibility of fabricating a superconductor using chemically doped 2D insulators is still limited.

We focused on hexagonal boron nitride (h-BN) as a candidate for the 2D superconductor. Bulk h-BN has a wide bandgap of approximately 6 eV, and it exists in the monolayer form. Although the application of h-BN as an inert and atomically flat insulating layer to realize novel 2D-material-based electronics[22–25] is widely investigated, its possibility as the superconductor had been ignored until recently. In our previously conducted study, we proposed that the Li-intercalated h-BN bilayer became a phonon-induced superconductor with Tc up to 25 K[26], thereby indicating that both the phonon properties and electron–phonon coupling (EPC) in doped h-BN are appropriate to realize superconductivity. However, the fabrication of superconducting Li-intercalated h-BN bilayer is experimentally challenging, as a different stacking order of the h-BN sheets than that of the original bulk structure is required to attain high Tc. Therefore, to find a promising candidate of an h-BN based superconductor with a more experimentally feasible structure than that of the bilayer structure, we investigate the properties of monolayer h-BN doped using each group-1 (Li, Na, K) and group-2 (Be, Mg, Ca, Sr, Ba) atom, respectively, by performing the ab-initio calculations of the electronic and phononic properties and EPC.

Consequently, we reveal that Sr- and Ba-doped monolayer h-BN and Ca-doped monolayer h-BN with 3.5% tensile strain are energetically stable and become superconductors with the Tc of approximately 5.83 K, 1.53 K, and 12.8 K, respectively. The strain in the Ca-doped case can be realized using the deformation of a polymer substrate or choosing a substrate on which the h-BN monolayer is fabricated. Furthermore, the Tc of the Ca-doped h-BN is higher than that of typical BCS superconductors such as Nb (9.29 K) and Pb (7.2 K). Therefore, the superconducting gap in the Ca-doped h-BN can be observed by performing transport measurement and/or low-temperature scanning tunneling spectroscopy[4,27]. Our results propose a new route of on-surface synthesis of 2D superconductors based on h-BN monolayer.

**Methods**

The ab-initio calculations were performed by using the Quantum-Espresso package[28,29]. The model structure of the doped monolayer h-BN is depicted in Fig. 1, wherein we assume the

$\sqrt{3} \times \sqrt{3}$R30° periodicity. Because the local-density approximation (LDA) for the exchange-correlation functional can satisfactorily reproduce the structure of h-BN[30,31], we used the norm-conserving pseudopotential by employing the Troullier–Martins method[32] with LDA. The plane-wave basis set was introduced, and the kinetic-energy cutoff and charge cutoff were set to 60 and 240 Ry, respectively. The Brillouin zone was sampled using a 24 × 24 × 1 k-mesh. The dynamical matrices and phonon deformation potentials were calculated using a Gamma-centered 12 × 12 × 1 q-mesh. In this study, we use the EPC matrix element that was directly obtained using Quantum-Espresso calculations, as the dispersive band structures above the Fermi level hampered the convergence in disentanglement and wannierization procedures to achieve high-precision Wannier interpolation[33]. To compensate the effect due to the sparse sampling of the Brillouin zone, we used a finer 120 × 120 × 1 k-mesh to interpolate the Fermi surface in the calculation of the Eliashberg function defined as follows:

$$\alpha^2 F(\omega) = \frac{1}{N_k N_q N(0)} \sum_{nk,mq,\nu} \left|g^{\nu}_{nk,mk+q}\right|^2 \delta(E_{nk}) \delta(E_{mk+q}) \delta(\omega - \omega^{\nu}_q),$$

where $N(0)$ denotes the density of states (DOS) at the Fermi level. $k$ and $q$ denote the wavevectors of electron and phonon, respectively. In addition, $N_k$ and $N_q$ denote the total number of k- and q-mesh grid points, respectively. Indexes $n$ and $m$ denote the electronic band index, $\nu$ is the phonon band index, and $\omega^{\nu}_q$ is the phonon energy. $E_{nk}$ and $E_{mk+q}$ denote the eigenvalues of the Kohn–Sham wavefunctions with respect to the Fermi level, and $g^{\nu}_{nk,mk+q}$ denotes the electron–phonon matrix element.

The superconducting transition temperature is estimated using the McMillan–Allen–Dynes formula[34,35] as follows:

$$T_c = \frac{\omega_{log}}{1.2} \exp\left[\frac{-1.04(1+\lambda)}{\lambda(1-0.62\mu^*)-\mu^*}\right]$$

where $\mu^*$ denotes the Coulomb repulsion pseudopotential, and we set $\mu^* = 0.14$. The EPC constant $\lambda$ is obtained by taking $\omega \to \infty$ in the following

$$\lambda(\omega) = 2\int_0^\omega d\omega' \frac{\alpha^2 F(\omega')}{\omega'}.$$

In addition, we decompose the EPC constant at the Fermi level with respect to the electron wavevector $k$ to determine the contribution of each electronic band. This $k$-resolved EPC is given as[10,36] follows:

$$\lambda_k = \sum_{q,\nu} \delta(\epsilon_{k+q}) \frac{|g^\nu_{nk,mk+q}|^2}{\omega_{q\nu}}.$$

**Results and Discussions**

First, we determined the stable structure in the doped h-BN and graphene by comparing the total energies with each other obtained by varying the values of the lattice constant, i.e., "a," and adsorbate–substrate distance, i.e., "h," (see Fig. 1). For the monolayer h-BN, no minimum was observed in its potential-energy curve as a function of h for Li, Na, K, Be, and Mg cases (see Fig. S1), meaning that the adsorption of these atoms was unstable. This point is further confirmed by the appearance of negative frequency in the phonon-dispersion curve (see Fig. S2). However, Ca, Sr, and Ba stably adsorbed on the h-BN monolayer. As depicted in Fig. 2 (a), in the Ba case, only a single minimum was observed in the potential-energy curve at approximately 5.0 a.u. Contrarily, in both the Ca and Sr cases, the potential-energy curve exhibited two minima, one at approximately 4.7 a.u. and the other at approximately 7.0 a.u. With equilibrium lattice constant, in the Sr case, the closer position of 4.66 a.u. was the global minimum; however, in the Ca case, the further position of 7.0 a.u. was the global minimum. To realize high Tc, a less distance between the adsorbate and h-BN sheet is favorable for enhancing the coupling between the low-frequency vibrational mode of the adsorbate and electronic state in h-BN. Therefore, we investigate the possibility of switching the global-minimum position by introducing tensile strain on the h-BN monolayer in the Ca-doped case. As shown in Fig. 2(b), the position of the global minimum depends on the lattice constant, i.e. strain. If the tensile strain is greater than 3.5%, the Ca-adatom favors the closer position of 4.66 a.u. The structure parameters, lattice constants, and distances between adsorbate and layer determined via the above calculations are summarized in Table SI. In addition, we show the parameters in Ca-, Sr-, and Ba-doped graphene used in the calculations for later comparison.

Subsequently, we evaluate the superconducting transition temperature for Sr- and Ba-doped h-BN with no strain and that for Ca-doped h-BN with tensile strain. The results of Tc are summarized in Table 1 together with the DOS at the Fermi level ($E_F$) and EPC constant ($\lambda$). Notably, our calculation results of Tc and $\lambda$ in Ca-doped graphene are slightly lower than those in a previously conducted report[8]. The differences between the calculation results is reasonable because compared with the previous report, we used a slightly larger lattice constant and distance between Ca atom and graphene layer.

As depicted in Fig. 3 and Table 1, doped h-BNs yield higher EPC constant and Tc than those of doped graphene systems (graphene is denoted as G). The higher EPC constant is mainly attributed to large $\alpha^2F(\omega)$ in the region of low-energy phonon, as seen in Figure 3. These phonon modes are localized at the adsorbate, and we highlighted them in the phonon-dispersion curves in Fig. 4. Both the in-plane (red) and out-of-plane (blue) vibrations of the adsorbate appear in the energy range of 0–30 meV. Although similar low-energy phonon modes by the adsorbate were also exhibited in the graphene cases, as compared with the modes in the h-BN cases, these modes contributed less to $\alpha^2F(\omega)$, as depicted in Fig. 3.

To clarify the difference between h-BN and graphene, we examine the electronic band structure depicted in Fig. 5. The projection of DOS from the adsorbate s-orbital onto the electronic-band structures indicates the presence of adsorbate bands that cross the Fermi level in all systems. These bands correspond to the interlayer state reported in previously conducted studies[8,15]. In doped graphene systems, the interlayer state is crucial for determining Tc. However, the similarity of the interlayer state between doped graphene and h-BN indicates that the Tc in these systems is determined by some other factor. Other than the interlayer states, several bands that originate from h-BN and graphene orbitals cross the Fermi level.

These band structures around the Fermi level result in Fermi surfaces with different shapes and positions, as depicted in Fig. 6. In both h-BN and graphene cases, the Fermi surface around the K point originates from the interlayer state, and those around the Γ point correspond to the band from h-BN or graphene states. The shape of the Fermi surfaces from the interlayer states resemble each another in both the doped h-BN and graphene cases, except for the size of the pocket. However, the shape of the inner Fermi surfaces around the Γ point is significantly different between the doped h-BN and graphene cases as follows. In the doped h-BN case, the inner Fermi surface has a snowflake-like shape, but it is hexagram-like in the doped graphene case. The projection of the k-resolved EPC ($\lambda_k$) shows that the electron–phonon interaction in the inner Fermi surface is crucial for determining Tc. Compared with doped graphenes, the doped h-BNs show greater $\lambda_k$ in the overall Brillouin zone. The most prominent enhancement of $\lambda_k$ is observed around the Γ point.

The enhancement of $\lambda_k$ around the Γ point in the Ca-doped h-BN is attributed to the spatial distribution of the wavefunction of the electronic states. In Fig. 7, we plot the charge-density distribution of each state that is indicated using crosses (× and +) in Fig. 5. The distinctive characteristics are seen in the state near the Γ point (marked by +) of the Ca-doped h-BN: the strong intensity of charge distribution connecting Ca and B atoms. Such a feature is retained but weakened at the states that are away from the Γ point and close to the M point (marked by ×). In the Ca-doped graphene case, the charge-distribution intensity between Ca and graphene layer is considerably smaller in both the states around the Γ and M points. Other than the interaction strength between Ca and substrate electronic state, symmetry is also important, as discussed in the following. Because of the symmetry difference in monolayer h-BN($C_{3v}$) and graphene($D_{6h}$), the number of possible rotation/reflection symmetry operations in the doped h-BN is fewer than that in the doped graphene. Therefore, the restriction on the coupling between intrinsic $\pi^*$–$\pi^*$ electrons mediated by the in-plane vibration of the Ca adsorbate is reduced in the h-BN case. This satisfactorily explains the coincidence of the peak position in $\alpha^2 F(\omega)$ and the projected phonon DOS onto the in-plane vibration of the Ca adsorbate, and the higher intensity of the peak of $\alpha^2 F(\omega)$ in the low-energy region in doped-h-BN than that in doped graphene. The above-mentioned symmetry restriction and coupling mechanism are different than those in the case of Li-doped graphene and other graphite intercalation compounds: the coupling between the interlayer state and $\pi^*$ electrons derived via the out-of-plane vibration of the dopant[8,11,14]. A similar superconductivity-enhancement mechanism via symmetry lowering was theoretically predicted in black phosphorene[37]. Therefore, instead of tuning the interlayer state, lowering the symmetry and introducing strong adsorbate–substrate interaction would be an alternative common strategy to increase the Tc of doped 2D semiconductors/insulators.

Finally, we discuss the 3.5% tensile strain in the Ca-doped h-BN case. Experimentally, high-quality h-BN thin films are fabricated via the exfoliation of bulk crystals or chemical vapor deposition on metal surfaces[38–41]. In the former case, strain can be applied via deforming the polymer substrate, and experiments with several percent strains up to 5.5% were reported[42,43]. However, in the latter case, the mismatch of the lattice constant between h-BN and substrate induces certain strain in the h-BN sheet. Notably, on Au and Ag substrates, the strain induced due to the lattice mismatch might be greater than 3.5%[44]. Therefore, the tensile strain of 3.5% on the h-BN monolayer might be experimentally possible.

**Conclusion**

We investigated the possibility of superconductivity in the monolayer h-BN that was doped using group-1 (Li, Na, K) and group-2 (Be, Mg, Ca, Sr, Ba) atoms by performing ab-initio calculations. We confirmed that each Ca, Sr, and Ba atom stably adsorbed on h-BN, while none of the other atoms (Li, Na, K, Be, Mg) did. In addition, in both the Ca and Sr cases, the potential-energy curve as a function of the adsorbate–surface distance exhibited two minima. Especially, in the Ca-doped case, the stable Ca-adsorption position with equilibrium lattice constant was far from the h-BN surface; however, a tensile strain greater than 3.5% enabled the adsorption of Ca atoms at a considerably closer distance from the h-BN surface. In all the Ca, Sr, and Ba cases, the electron doping made h-BN metallic. Furthermore, we calculated the Tc of both Sr- and Ba-doped h-BN with equilibrium lattice constant and that of Ca-doped h-BN with the tensile strain of 3.5%. The obtained Tc values were 5.83 K, 1.53 K, and 12.8 K, respectively. All these Tc values were considerably higher than those in the doped-graphene case. A significant difference between doped h-BN and graphene appears in the k-resolved EPC that is projected on the Fermi surface. In the h-BN case, the intrinsic $\pi^*$ orbitals formed a Fermi surface around the $\Gamma$ point. The EPC in this $\pi^*$ orbitals mediated by the adsorbate in-plane vibration was enhanced because of the strong interaction between Ca and B atoms and lower symmetry in h-BN compared to graphene, thereby explaining the origin of high Tc in Ca-doped h-BN.

**Acknowledgments**

The following financial support is acknowledged: JST PRESTO JPMJPR17I7 (E. M.), and MEXT KAKENHI 17H05330 (S. W. and E. M.). The calculations were performed using the following computer facilities: Institute of Solid State Physics (Kashiwa, Japan), Information Technology Center (University of Tokyo, Tokyo, Japan), Center for Computational Materials Science (Tohoku University, Sendai, Japan), and Research Center for Computational Science (Okazaki, Japan).

**References**

(1) Bhimanapati, G. R.; Lin, Z.; Meunier, V.; Jung, Y.; Cha, J.; Das, S.; Xiao, D.; Son, Y.; Strano, M. S.; Cooper, V. R.; et al. Recent Advances in Two-Dimensional Materials beyond Graphene. *ACS Nano*. **2015**, 9, 11509–11539.

(2) Uchihashi, T. Two-Dimensional Superconductors with Atomic-Scale Thickness. *Supercond. Sci. Technol.* **2017**, *30*, 013002.

(3) Matetskiy, A. V.; Ichinokura, S.; Bondarenko, L. V.; Tupchaya, A. Y.; Gruznev, D. V.;

Zotov, A. V.; Saranin, A. A.; Hobara, R.; Takayama, A.; Hasegawa, S. Two-Dimensional Superconductor with a Giant Rashba Effect: One-Atom-Layer Tl-Pb Compound on Si(111). *Phys. Rev. Lett.* **2015**, *115,* 147003.

(4) Eom, D.; Qin, S.; Chou, M. Y.; Shih, C. K. Persistent Superconductivity in Ultrathin Pb Films: A Scanning Tunneling Spectroscopy Study. *Phys. Rev. Lett.* **2006**, *96*, 027005.

(5) Liu, D.; Zhang, W.; Mou, D.; He, J.; Ou, Y. B.; Wang, Q. Y.; Li, Z.; Wang, L.; Zhao, L.; He, S.; et al. Electronic Origin of High-Temperature Superconductivity in Single-Layer FeSe Superconductor. *Nat. Commun.* **2012**, *3*, 1–6.

(6) Ge, J. F.; Liu, Z. L.; Liu, C.; Gao, C. L.; Qian, D.; Xue, Q. K.; Liu, Y.; Jia, J. F. Superconductivity above 100 K in Single-Layer FeSe Films on Doped $SrTiO_3$. *Nat. Mater.* **2015**, *14*, 285–289.

(7) Xi, X.; Wang, Z.; Zhao, W.; Park, J. H.; Law, K. T.; Berger, H.; Forró, L.; Shan, J.; Mak, K. F. Ising Pairing in Superconducting NbSe2 Atomic Layers. *Nat. Phys.* **2016**, *12* , 139–143.

(8) Profeta, G.; Calandra, M.; Mauri, F. Phonon-Mediated Superconductivity in Graphene by Lithium Deposition. *Nat. Phys.* **2012**, *8*, 131–134.

(9) Ludbrook, B. M.; Levy, G.; Nigge, P.; Zonno, M.; Schneider, M.; Dvorak, D. J.; Veenstra, C. N.; Zhdanovich, S.; Wong, D.; Dosanjh, P.; et al. Evidence for Superconductivity in Li-Decorated Monolayer Graphene. *Proc. Natl. Acad. Sci.* **2015**, *112* , 11795–11799.

(10) Margine, E. R.; Giustino, F. Two-Gap Superconductivity in Heavily n -Doped Graphene: Ab Initio Migdal-Eliashberg Theory. *Phys. Rev. B* **2014**, *90*, 014518.

(11) Margine, E. R.; Lambert, H.; Giustino, F. Electron-Phonon Interaction and Pairing Mechanism in Superconducting Ca-Intercalated Bilayer Graphene. *Sci. Rep.* **2016**, *6*,

21414.

(12) Ichinokura, S.; Sugawara, K.; Takayama, A.; Takahashi, T.; Hasegawa, S. Superconducting Calcium-Intercalated Bilayer Graphene. *ACS Nano* **2016**, *10*, 2761–2765.

(13) Kanetani, K.; Sugawara, K.; Sato, T.; Shimizu, R.; Iwaya, K.; Hitosugi, T.; Takahashi, T. Ca Intercalated Bilayer Graphene as a Thinnest Limit of Superconducting $C_6Ca$. *Proc. Natl. Acad. Sci.* **2012**, *109*, 19610–19613.

(14) Yang, S. L.; Sobota, J. A.; Howard, C. A.; Pickard, C. J.; Hashimoto, M.; Lu, D. H.; Mo, S. K.; Kirchmann, P. S.; Shen, Z. X. Superconducting Graphene Sheets in $CaC_6$ Enabled by Phonon-Mediated Interband Interactions. *Nat. Commun.* **2014**, *5*, 4493.

(15) Zheng, J.-J.; Margine, E. R. First-Principles Calculations of the Superconducting Properties in Li-Decorated Monolayer Graphene within the Anisotropic Migdal-Eliashberg Formalism. *Phys. Rev. B* **2016**, *94*, 064509.

(16) Ye, J. T.; Zhang, Y. J.; Akashi, R.; Bahramy, M. S.; Arita, R.; Iwasa, Y. Superconducting Dome in a Gate-Tuned Band Insulator. *Science* **2012**, *338,* 1193–1196.

(17) Saito, Y.; Nakamura, Y.; Bahramy, M. S.; Kohama, Y.; Ye, J.; Kasahara, Y.; Nakagawa, Y.; Onga, M.; Tokunaga, M.; Nojima, T.; et al. Superconductivity Protected by Spin-Valley Locking in Ion-Gated MoS2. *Nat. Phys.* **2016**, *12*, 144–149.

(18) Costanzo, D.; Jo, S.; Berger, H.; Morpurgo, A. F. Gate-Induced Superconductivity in Atomically Thin $MoS_2$ Crystals. *Nat. Nanotechnol.* **2016**, *11*, 339–344.

(19) Saito, Y.; Kasahara, Y.; Ye, J.; Iwasa, Y.; Nojima, T. Metallic Ground State in an Ion-Gated Two-Dimensional Superconductor. *Science.* **2015**, *350*, 409–413.

(20) Zhang, R.; Tsai, I. L.; Chapman, J.; Khestanova, E.; Waters, J.; Grigorieva, I. V. Superconductivity in Potassium-Doped Metallic Polymorphs of $MoS_2$. *Nano Lett.* **2016**,

*16*, 629–636.

(21) Zhang, R.; Waters, J.; Geim, A. K.; Grigorieva, I. V. Intercalant-Independent Transition Temperature in Superconducting Black Phosphorus. *Nat. Commun.* **2017**, *8*, 15036.

(22) Dean, C. R.; Young, A. F.; Meric, I.; Lee, C.; Wang, L.; Sorgenfrei, S.; Watanabe, K.; Taniguchi, T.; Kim, P.; Shepard, K. L.; et al. Boron Nitride Substrates for High-Quality Graphene Electronics. *Nat. Nanotechnol.* **2010**, *5*, 722–726.

(23) Liu, Z.; Ma, L.; Shi, G.; Zhou, W.; Gong, Y.; Lei, S.; Yang, X.; Zhang, J.; Yu, J.; Hackenberg, K. P.; et al. In-Plane Heterostructures of Graphene and Hexagonal Boron Nitride with Controlled Domain Sizes. *Nat. Nanotechnol.* **2013**, *8*, 119–124.

(24) Wang, J.; Yao, Q.; Huang, C. W.; Zou, X.; Liao, L.; Chen, S.; Fan, Z.; Zhang, K.; Wu, W.; Xiao, X.; et al. High Mobility $MoS_2$ Transistor with Low Schottky Barrier Contact by Using Atomic Thick H-BN as a Tunneling Layer. *Adv. Mater.* **2016**, *28*, 8302–8308.

(25) Kubota, Y.; Watanabe, K.; Tsuda, O.; Taniguchi, T. Deep Ultraviolet Light-Emitting Hexagonal Boron Nitride Synthesized at Atmospheric Pressure. *Science* **2007**, *317*, 932–934.

(26) Shimada, N. H.; Minamitani, E.; Watanabe, S. Theoretical Prediction of Phonon-Mediated Superconductivity with Tc ≈ 25K in Li-Intercalated Hexagonal Boron Nitride Bilayer. *Appl. Phys. Express* **2017**, *10*, 093101.

(27) Pan, S. H.; Hudson, E. W.; Davis, J. C. Vacuum Tunneling of Superconducting Quasiparticles from Atomically Sharp Scanning Tunneling Microscope Tips. *Appl. Phys. Lett.* **1998**, *73*, 2992–2994.

(28) Giannozzi, P.; Andreussi, O.; Brumme, T.; Bunau, O.; Buongiorno Nardelli, M.; Calandra, M.; Car, R.; Cavazzoni, C.; Ceresoli, D.; Cococcioni, M.; et al. Advanced Capabilities for Materials Modelling with Quantum ESPRESSO. *J. Phys. Condens.*

*Matter* **2017**, *29*, 465901.

(29) Giannozzi, P.; Baroni, S.; Bonini, N.; Calandra, M.; Car, R.; Cavazzoni, C.; Ceresoli, D.; Chiarotti, G. L.; Cococcioni, M.; Dabo, I.; et al. QUANTUM ESPRESSO: A Modular and Open-Source Software Project for Quantum Simulations of Materials. *J. Phys. Condens. Matter* **2009**, *21*, 395502.

(30) Kern, G.; Kresse, G.; Hafner, J. Ab Initio Calculation of the Lattice Dynamics and Phase Diagram of Boron Nitride. *Phys. Rev. B* **1999**, *59*, 8551–8559.

(31) Hamdi, I.; Meskini, N. Ab Initio Study of the Structural, Elastic, Vibrational and Thermodynamic Properties of the Hexagonal Boron Nitride: Performance of LDA and GGA. *Physica. B* **2010**, 405, 2785-2794.

(32) Troullier, N.; Martins, J. L. Efficient Pseudopotentials for Plane-Wave Calculations. *Phys. Rev. B* **1991**, *43*, 1993–2006.

(33) Giustino, F.; Cohen, M. L.; Louie, S. G. Electron-Phonon Interaction Using Wannier Functions. *Phys. Rev. B* **2007**, *76*, 165108.

(34) McMillan, W. L. Transition Temperature of Strong-Coupled Superconductors. *Phys. Rev.* **1968**, *167*, 331–344.

(35) Dynes, R. C. McMillan's Equation and the Tc of Superconductors. *Solid State Commun.* **1972**, *10*, 615–618.

(36) Margine, E. R.; Giustino, F. Anisotropic Migdal-Eliashberg Theory Using Wannier Functions. *Phys. Rev. B* **2013**, *87*, 024505.

(37) Huang, G. Q.; Xing, Z. W.; Xing, D. Y. Prediction of Superconductivity in Li-Intercalated Bilayer Phosphorene. *Appl. Phys. Lett.* **2015**, *106*, 113107.

(38) Sutter, P.; Lahiri, J.; Albrecht, P.; Sutter, E. Chemical Vapor Deposition and Etching of High-Quality Monolayer Hexagonal Boron Nitride Films. *ACS Nano* **2011**, *5*, 7303–

7309.

(39) Jin, C.; Lin, F.; Suenaga, K.; Iijima, S. Fabrication of a Freestanding Boron Nitride Single Layer and Its Defect Assignments. *Phys. Rev. Lett.* **2009**, *102*, 195505.

(40) Gao, Y.; Ren, W.; Ma, T.; Liu, Z.; Zhang, Y.; Liu, W. Bin; Ma, L. P.; Ma, X.; Cheng, H. M. Repeated and Controlled Growth of Monolayer, Bilayer and Few-Layer Hexagonal Boron Nitride on Pt Foils. *ACS Nano* **2013**, *7*, 5199–5206.

(41) Auwärter, W. Hexagonal Boron Nitride Monolayers on Metal Supports: Versatile Templates for Atoms, Molecules and Nanostructures. *Surface Science Reports*. **2019**, 74, 1-95.

(42) Mendelson, N.; Doherty, M.; Toth, M.; Aharnovish, I.; Tran, T. T. Tuning of Quantum Emitters in Hexagonal Boron Nitride. *arXiv:1911.08072* **2019**. https://doi.org/http://arxiv.org/abs/1911.08072.

(43) Androulidakis, C.; Koukaras, E. N.; Poss, M.; Papagelis, K.; Galiotis, C.; Tawfick, S. Strained Hexagonal Boron Nitride: Phonon Shift and Grüneisen Parameter. *Phys. Rev. B* **2018**, *97*, 241414.

(44) Laskowski, R.; Blaha, P.; Schwarz, K. Bonding of Hexagonal BN to Transition Metal Surfaces: An Ab Initio Density-Functional Theory Study. *Phys. Rev. B* **2008**, *78*, 045409.

Figures

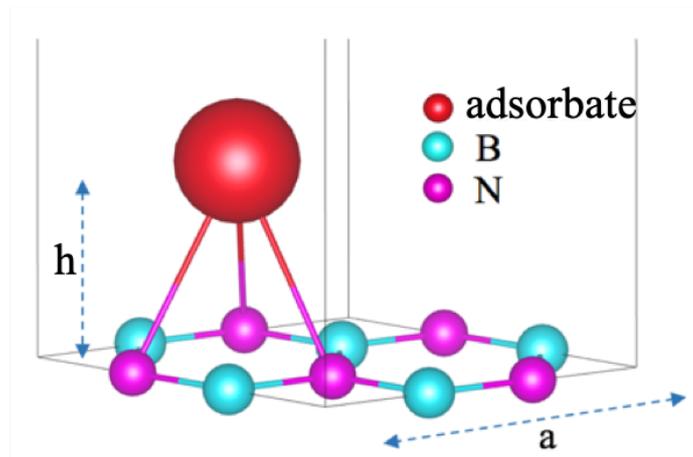

Figure 1 Structure of chemically doped h-BN.

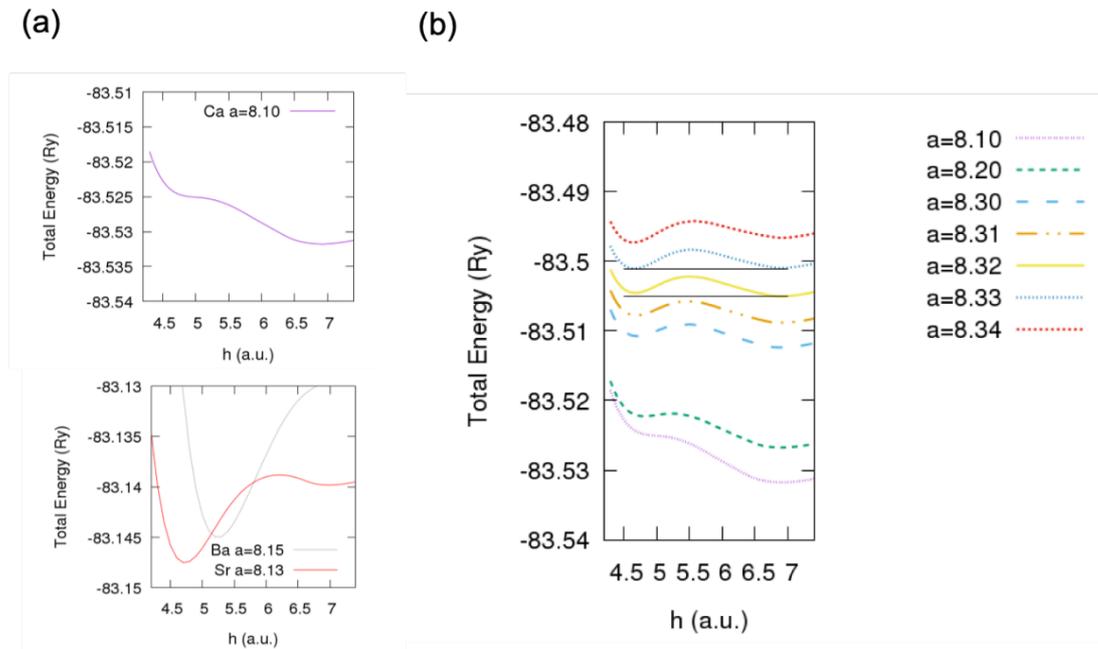

Figure 2

(a) Potential-energy curve as a function of the adsorbate–surface distance for the Ca-, Ba-, and Sr-doped cases. (b) Dependence of the potential-energy curve in the Ca-doped case on the lattice constant of h-BN.

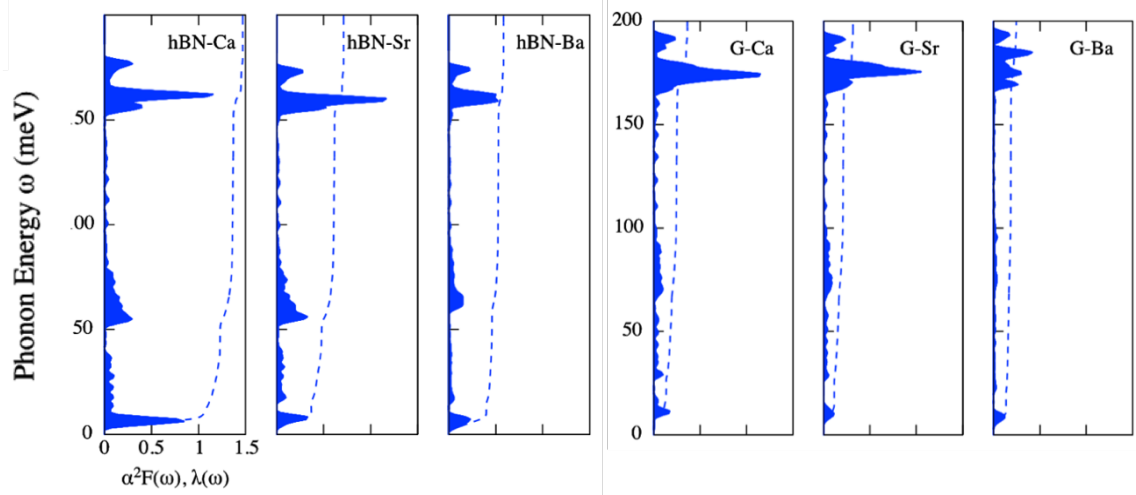

Figure 3 $\alpha^2F(\omega)$ and $\lambda(\omega)$ of doped h-BN and graphene. Notably, $\alpha^2F(\omega)$ and $\lambda(\omega)$ are represented using the blue region and dotted line, respectively.

Table 1 Summary of the DOS at the Fermi level (N($E_F$)), EPC ($\lambda$), and Tc.

| composition | N($E_F$) | $\lambda$ | Tc (K) |
|---|---|---|---|
| hBN-Ca | 2.36 | 1.45 | 12.8 |
| hBN-Sr | 2.37 | 0.71 | 5.83 |
| hBN-Ba | 2.14 | 0.57 | 1.53 |
| G-Ca | 1.27 | 0.36 | 0.26 |
| G-Sr | 1.20 | 0.31 | 0.04 |
| G-Ba | 1.14 | 0.24 | 0.00 |

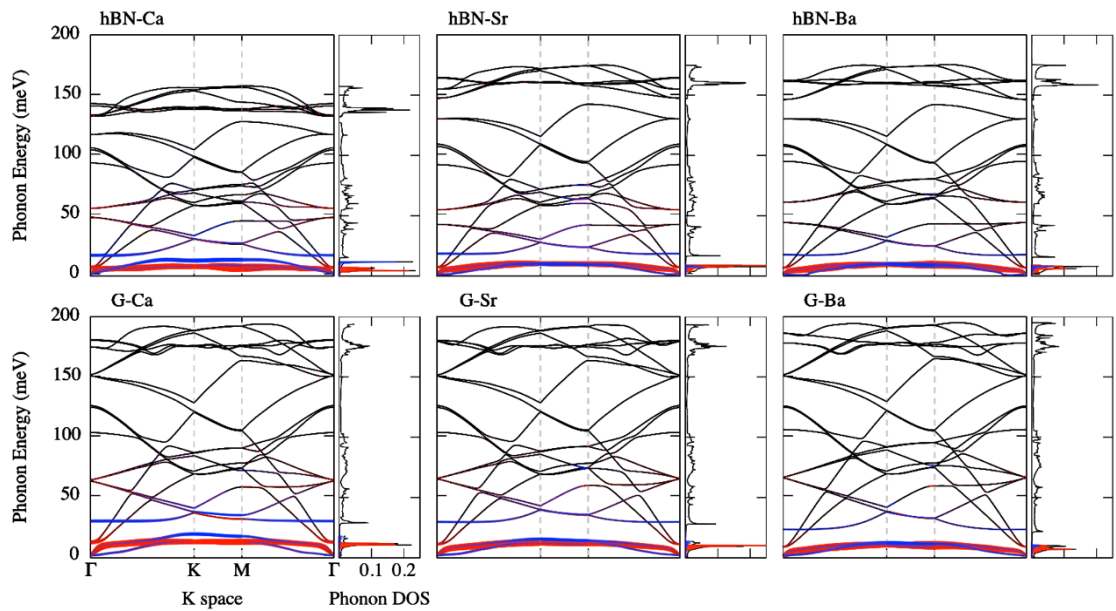

Figure 4 Phonon band dispersion of doped h-BN and graphene. The red (blue) lines in bands and phonon DOS represent the contribution of the adsorbate in-plane (out-of-plane) vibrational modes.

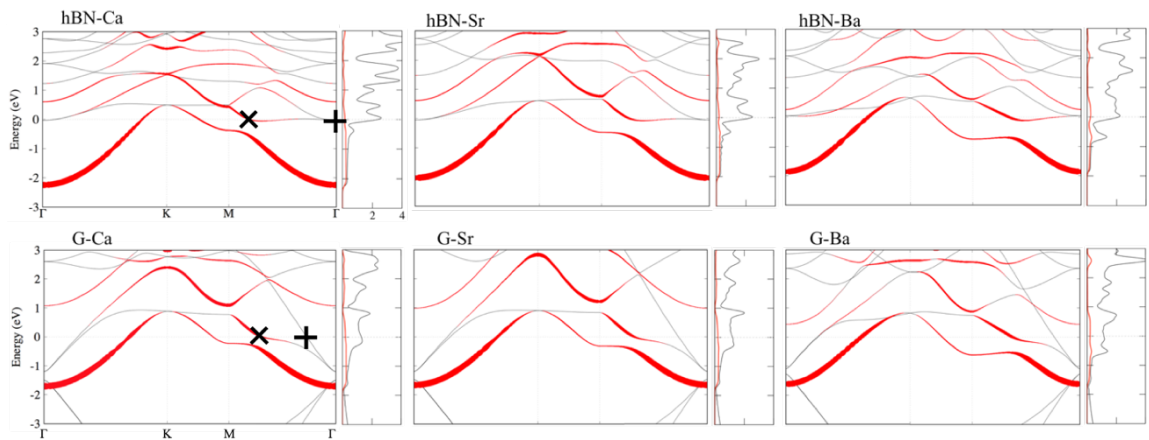

Figure 5 Electronic band dispersions and DOS for doped h-BN and graphene. The thickness of the red lines in band dispersion is proportional to the contribution of the adsorbate s-orbital. The unit of DOS is states/eV/spin. In addition, × and + represent the eigenstate for which the charge distribution in Figure 7 is plotted.

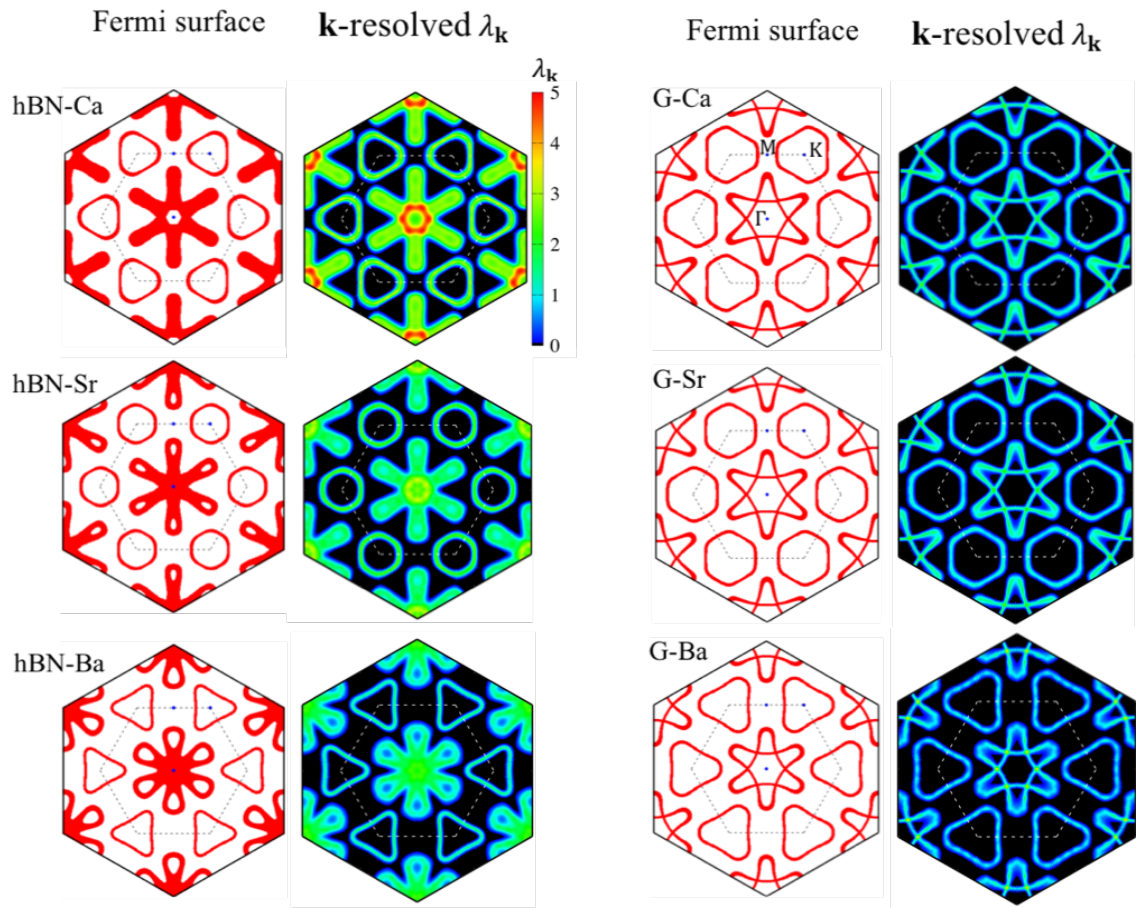

Figure 6 Fermi surfaces and the k-resolved EPC projected on the Fermi surface in doped h-BN and graphene.

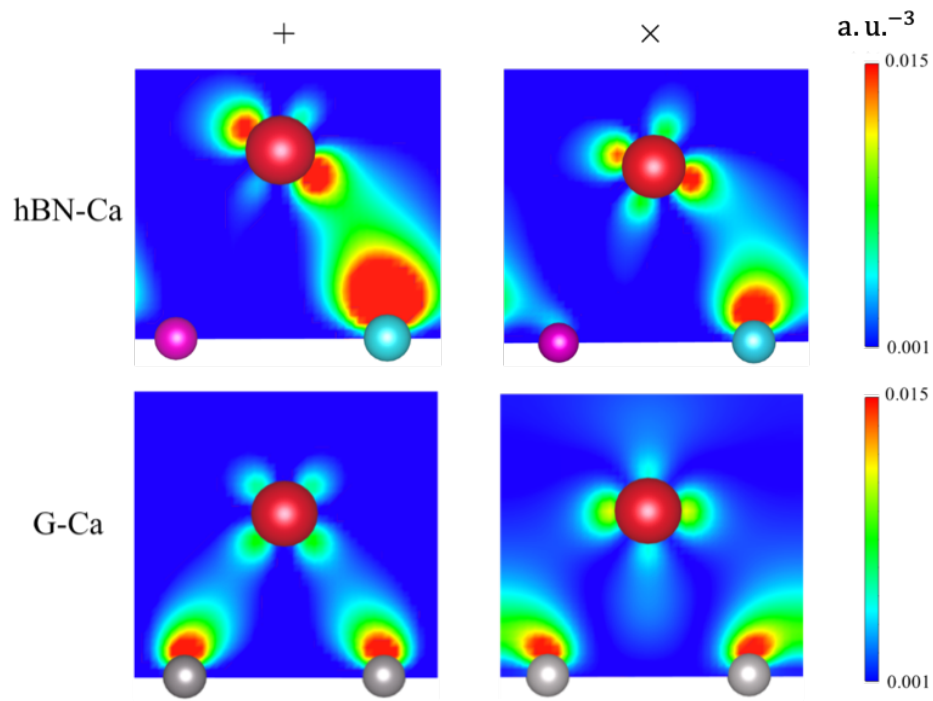

Figure 7 Charge-density distributions of the eigenstate (specified by × and + in Fig. 5) for Ca-doped graphene and h-BN. The charge densities (a.u.$^{-3}$) are normalized so that their integrated value over a unit cell is 1.